\documentclass[preprint,12pt]{elsarticle}
\usepackage{graphicx}
\usepackage{amsfonts,amsmath,amssymb,bm}
\usepackage[dvips,usenames]{color}

\newcommand{\eg}{{\it e.g.}}
\newcommand{\ie}{{\it i.e.}}

\journal{Physica A}

\begin{document}

\begin{frontmatter}



\title{Proportionate vs disproportionate distribution of wealth of two individuals in a tempered Paretian ensemble}


\author[label1]{G. Oshanin}
\address[label1]{Laboratoire de Physique Th{\'e}orique de la Mati{\`e}re
Condens{\'e}e (UMR CNRS 7600), Universit{\'e} Pierre et Marie Curie (Paris 6) -
4 Place Jussieu, 75252 Paris, France}

\author[label2]{Yu. Holovatch}
\address[label2]{Institute for Condensed Matter Physics,
National Academy of Sciences of Ukraine,
1 Svientsitskii Street, Lviv, 79011 Ukraine}

\author[label3]{G. Schehr}
\address[label3]{Laboratoire de Physique Th\'eorique, Universit\'e de Paris-Sud, France}

\begin{abstract}
We study the distribution $P(\omega)$ of the random variable $\omega = x_1/(x_1 + x_2)$, where $x_1$ and $x_2$ are  the wealths of two individuals selected at random from
 the same tempered Paretian ensemble
 characterized by the distribution
$\Psi(x) \sim \phi(x)/x^{1 + \alpha}$,
where $\alpha > 0$ is the Pareto index and $\phi(x)$ is the cut-off function.
We consider two forms of $\phi(x)$:
a bounded function $\phi(x) = 1$ for $L \leq x \leq H$, and zero otherwise,
and a smooth
exponential function $\phi(x) = \exp(-L/x - x/H)$.
In both cases $\Psi(x)$ has moments of arbitrary order.
 We show that, for $\alpha > 1$,  $P(\omega)$
always has a unimodal form and is peaked at $\omega = 1/2$, so that
most probably $x_1 \approx x_2$.
For $0 < \alpha < 1$ we observe a more complicated behavior which depends on the value of $\delta = L/H$. In particular, for
$\delta < \delta_c$ - a certain threshold value - $P(\omega)$ has a three-modal (for a bounded $\phi(x)$) and a bimodal $M$-shape (for an exponential $\phi(x)$) form which signifies
 that in such ensembles the wealths $x_1$ and $x_2$ are disproportionately different.
\end{abstract}


\end{frontmatter}

\section{Introduction}

At the close of the nineteenth century
the Italian economist Vilfredo Pareto,
while studying the statistics of human income and wealth, discovered
that the distribution $\Psi(x)$ of both
has a remarkable power-law form, $\Psi(x) \sim A/x^{1 + \alpha}$  \cite{pareto},
where  $\alpha > 0$ is a parameter which
is now called the Pareto index.
Thereafter, Pareto's law was verified for various countries, both for the wealth of individuals and for their income,
and
was also observed in empirical
data
for diverse scientific fields
(see, \eg, Refs.~\cite{newman,mit,levy,iddo,amit} and references therein).
It was realized that the Pareto index measured from the income
distribution is typically larger than the index $\alpha$ deduced from the distribution of
wealth. This is, of course, consistent with the general observation that in market economies wealth is more unequally distributed than income \cite{samuelson}. For the wealth distribution,
the observed Pareto index $\alpha$ is as high as $2.3-2.5$ for developed countries,
but may be as low as $0.81$ for a developing economy like India \cite{sinha}.

As a matter of fact, only very few real-world distributions may
follow a power law over their entire range.
As a distribution of wealth, which is large but nonetheless finite, the Pareto distribution (PD)
is not an exception.
In the analysis of data, one uses different forms of truncated PDs, such as, \eg,
the bounded distribution
\begin{equation}
\label{trunc1}
\Psi(x) = \frac{\alpha L^{\alpha}}{1 - \delta^{\alpha}} \,
\begin{cases}
1/x^{1 + \alpha}\,,~\mathrm{for}~L \leq x \leq H\,,\\
0\,,~\mathrm{otherwise}\,,
\end{cases}
\end{equation}
where $L$ and $H$ are lower and upper cut-offs, respectively, and $\delta = L/H < 1$.
On the other hand,
some real-world examples are not that abrupt so that one seeks to fit the data using a
smoother truncation procedure, \eg,
\begin{equation}
\label{trunc2}
 \Psi(x) = \frac{1}{2}  \frac{\left(L H\right)^{\alpha/2}}{K_{\alpha}(2 \sqrt{\delta})} \frac{1}{x^{1 + \alpha}} \exp\left(- \frac{L}{x} - \frac{x}{H}\right) \;,
\end{equation}
where $K_{\alpha}(x)$ is the modified Bessel function.
In both Eqs.~(\ref{trunc1}) and (\ref{trunc2}) a power-law emerges as an intermediate behavior,
so that $\Psi(x)$ has moments of arbitrary order.

Before we proceed, it might be useful to remark that the exponentially-truncated distribution in Eq.~(\ref{trunc2}) is rather ubiquitous and appears in many areas in physics. To name but a few
we mention the distribution of the first passage times for random motion with a bias \cite{red}, the distribution of times between action potentials (or the ISI distribution)
in the integrate-and-fire model of neuron dynamics \cite{man}, the  distribution of the stopping distances for the sliding motion of a solid block on an inclined heterogeneous plane
\cite{lima}, the avalanche life-time distribution in the mean-field version of the Bak-Sneppen model \cite{flyv}, probability current distribution in disordered systems \cite{current} and the distribution of the number of times a particle diffusing in a sphere hits its boundary \cite{hit}.
Thus our subsequent analysis applies to these systems as well.

In this paper we seek an answer to the following question:
Suppose  in a given tempered Paretian ensemble one selects at random
 two individuals with wealths $x_1$ and $x_2$, respectively.
 How different are $x_1$ and $x_2$? In quest for the answer, we introduce
 a random variable,
\begin{equation}
\label{omega}
\omega = \frac{x_1}{x_1 + x_2},
\end{equation}
which defines the contribution of
one of two individuals
to the total wealth $x_1 + x_2$ of two of them, and calculate its distribution function $P(\omega)$.

Note  that the random variables in Eq.~(\ref{omega}) have apparently
first been studied
 in Ref.~\cite{iddo1} within a general
context of
heavy-tailed, non-truncated distributions not
 having a second, nor even a first, moment.
It was realized that for $\alpha < 1$ the distribution $P(\omega)$ has a characteristic $U$-shape form with a minimum at $\omega = 1/2$. In our language, this means  that for such unbounded distributions $x_1 = x_2$ is the least probable event, and the wealth distribution is highly disproportionate. Note that recently the distribution of more general variables of the form $x_1/(x_1 + x_2 + \ldots + x_N)$ have been studied in Ref.~\cite{we1} for different heavy-tailed and tempered parental distributions $\Psi(x)$.

In our case,
all moments of $\Psi(x)$ exist. Moreover, since $\langle x_1^n \rangle \equiv \langle x_2^n \rangle$ for arbitrary $n$, one may generally
expect that $P(\omega)$ will be a unimodal distribution peaked at $\omega = 1/2$. We set out to show instead that this is not always the case but that surprisingly, the distribution
 $P(\omega)$ exhibits a rather rich,
 sometimes counterintuitive behavior. In particular, we will demonstrate that
 for tempered distributions with $0 < \alpha < 1$ the distribution $P(\omega)$ undergoes a transition
 from a bell-shaped form (so that $x_1 = x_2$ is the most probable event)
   to an $M$-shaped (for smoothly truncated $\Psi(x)$) and a three-modal (for bounded $\Psi(x)$) forms when $\delta = L/H$ becomes less than some threshold value $\delta_c$. This furnishes another striking example that random variables with tempered heavy tails can be similar, in important respects, to random variables with non-truncated heavy tails \cite{sam}.

\section{The distribution $P(\omega)$: A general result}

Let $\Phi(\lambda)$ denote the moment generating function of $\omega$:
\begin{equation}
\Phi(\lambda) = \int_0^{\infty} \int_0^{\infty} dx_1 dx_2 \Psi(x_1) \Psi(x_2) \exp\left(- \lambda \frac{x_1}{x_1 + x_2} \right).
\end{equation}
Performing the integration over $dx_1$, we formally change the integration variable $x_1 \to \omega$, so that
\begin{eqnarray}
\Phi(\lambda) &=& \int_0^{1} \frac{d\omega}{(1 - w)^2} \exp\left(- \lambda \omega\right) \nonumber\\ &\times& \int_0^{\infty} x_2 \, dx_2 \, \Psi\left(\frac{\omega}{1 - \omega} x_2\right) \, \Psi(x_2),
\end{eqnarray}
from which we can read off the desired distribution:
\begin{equation}
\label{dist}
P(\omega) = \frac{1}{(1 - \omega)^2} \int_0^{\infty} x \, dx \,  \Psi(\frac{\omega}{1 - \omega} x) \,
\Psi(x).
\end{equation}
One may readily verify that $P(\omega)$ is normalized, \ie, $\int^1_0 d\omega P(\omega) \equiv 1$, once $\Psi(x)$ is normalized. Note, as well,
that for arbitrary $\Psi(x)$ one finds from Eq.~(\ref{dist}) that $\langle \omega \rangle \equiv \int^1_0 \omega \, d\omega P(\omega) = 1/2$, which can be readily seen from the following simple argument:
\begin{eqnarray}
1 = \big<\frac{x_1 + x_2}{x_1 + x_2}\big> = \big<\frac{x_1}{x_1 + x_2}\big> + \big<\frac{x_2}{x_1 + x_2}\big> = 2 \big<\frac{x_1}{x_1 + x_2}\big>.
\end{eqnarray}
We will show in what follows that the average behavior is not representative, and under certain conditions
does not coincide with typical or most probable behavior.

\begin{figure}[b]
  \centerline{\includegraphics*[width=0.65\textwidth]{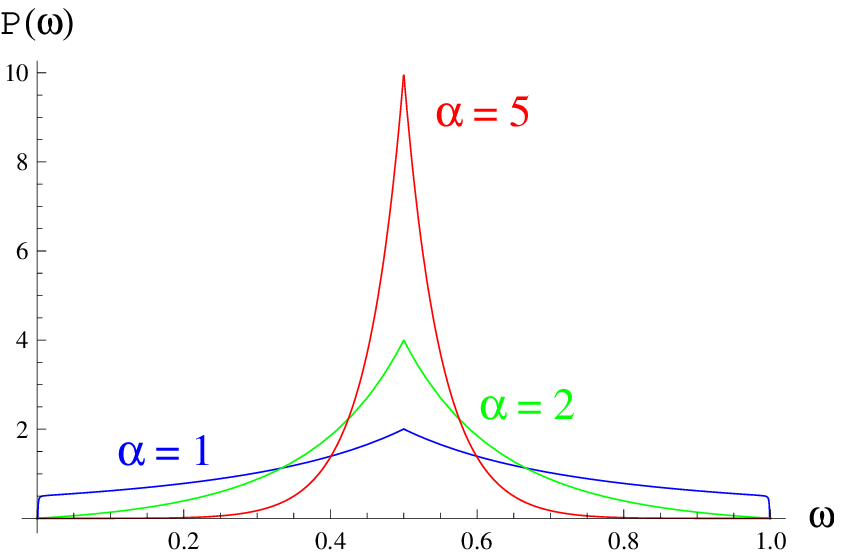}}
\caption{$P(\omega)$ in Eq.~(\ref{expl}) for $\delta = 0.001$ and
different
$\alpha$.
}
  \label{sketch2}
\end{figure}

\section{The distribution $P(\omega)$ for a bounded Pareto law}

Consider first the bounded PD in Eq.~(\ref{trunc1}).
Note that
for such a distribution,
the choice of a random variable as in Eq.~(\ref{omega})
will automatically lead to a
symmetric distribution function that depends on $L$ and $H$ only via the ratio $\delta = L/H$, and
has a support not on $[0,1]$ but on a smaller interval $[\omega_c,1-\omega_c]$
where,
 for $L \leq x \leq H$, $\omega_c=\delta/(1+\delta)$.
Substituting Eq.~(\ref{trunc1}) into Eq.~(\ref{dist})  we find
\begin{eqnarray}
\label{expl}
P(\omega) &=& \frac{\alpha}{2 (1 - \delta^{\alpha})^2} \, \frac{1}{\omega^{1 + \alpha} (1 - \omega)^{1 + \alpha}}  \nonumber\\
&\times&
\begin{cases}
{\rm M}^{- 2 \alpha}(\omega) - \delta^{2 \alpha} {\rm m}^{- 2 \alpha}(\omega)\,,~\mathrm{for}~\omega_c \leq \omega \leq 1 - \omega_c\,,\\
0\,,~\mathrm{for}~\omega < \omega_c\,~\mathrm{or}~\,\omega > 1 - \omega_c~~,
\end{cases}
\end{eqnarray}
with
\begin{eqnarray}
{\rm m}(\omega) &=& {\rm min}\left(\frac{1}{\omega},\frac{1}{1 - \omega}\right), \nonumber\\
{\rm M}(\omega) &=& {\rm max}\left(\frac{1}{\omega},\frac{1}{1 - \omega}\right).
\end{eqnarray}
A straightforward analysis shows that for $\alpha \geq 1$, regardless of the value of $\delta$,
$P(\omega)$ is unimodal with a maximum at $\omega = 1/2$. This signifies
that in this case most probably $x_1 = x_2$, and hence,
both individuals in such a Paretian ensemble
 contribute
to the total wealth \textit{proportionally}.

\begin{figure}[b]
  \centerline{\includegraphics*[width=0.65\textwidth]{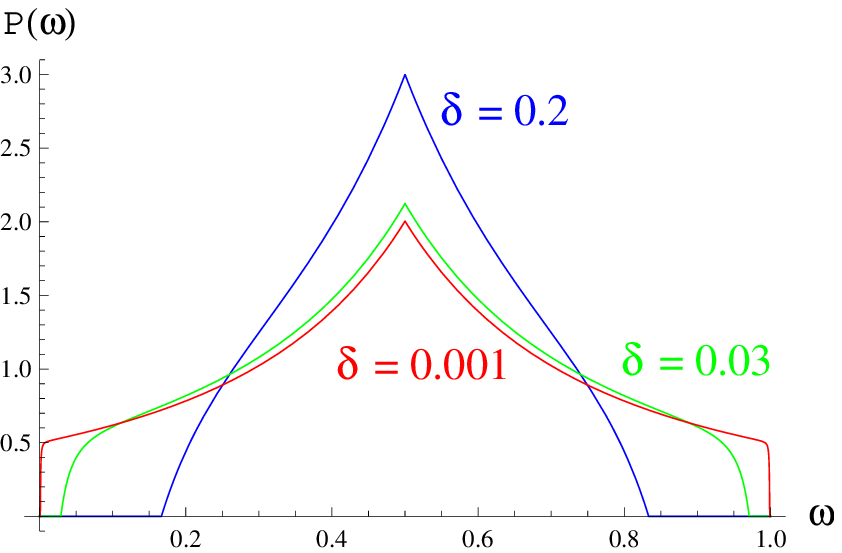}}
\caption{$P(\omega)$ in Eq.~(\ref{expl}) for
$\alpha = 1$ and
different values of
$\delta$.
}
  \label{sketch1}
\end{figure}

The distribution $P(\omega)$ for $\alpha \geq 1$ is depicted in Figs.~\ref{sketch2} and \ref{sketch1}.
It has a cusp-like maximum at $\omega = 1/2$;
$P(\omega)$ becomes narrower and the height of the maximum, $P(\omega=1/2)$, increases as
the Pareto index $\alpha$ increases. On the other hand, $P(\omega)$ broadens as $\delta \to 0$ and $P(\omega=1/2)$ decreases.
This is, of course, quite a plausible behavior which
one may expect on intuitive grounds.

\begin{figure}[ht]
  \centerline{\includegraphics[width=0.65\textwidth]{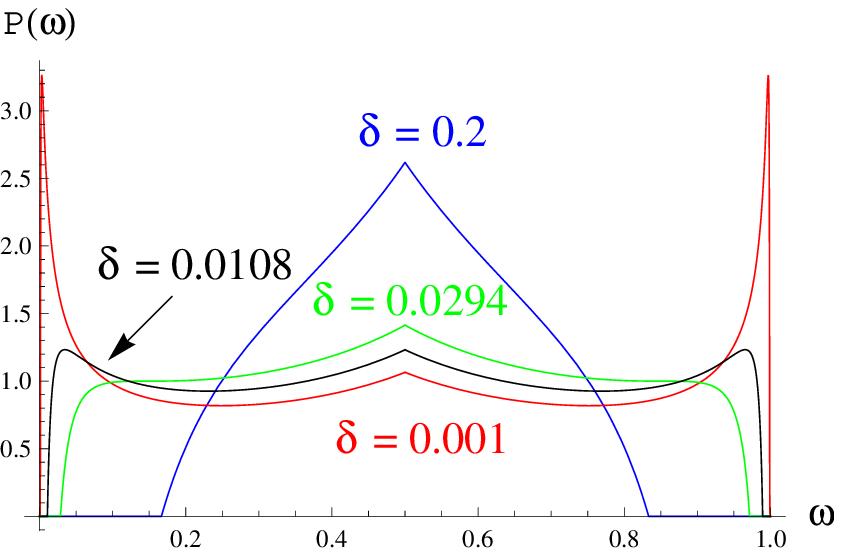}}
\caption{$P(\omega)$ in Eq.~(\ref{expl}) for $\alpha = 1/2$ and  different $\delta$.
}
  \label{sketch3}
\end{figure}

 When $\alpha < 1$ the situation appears to be more interesting and complicated.
 As in the previous case, here $P(\omega)$ always has a cusp-like maximum at $\omega = 1/2$.
However, there exists a critical value $\delta_c$ of
the parameter $\delta$ at which two inflection points emerge symmetrically in the regions
$\omega < 1/2$ and $\omega>1/2$.
For $\delta < \delta_c$ each of these inflection points splits into a minimum and a maximum
 so that $P(\omega)$ attains a three-modal, $W$-shaped form.
 When $\delta$ becomes yet smaller than some other critical
 $\delta_{cc}$, one observes that the value of $P(\omega)$ at these local maxima
  becomes greater than $P(\omega = 1/2)$. This signifies that in such a Paretian ensemble
  most probably the two individuals, selected at random, have disproportionate wealth
  since $x_1 \gg x_2$ or $x_1 \ll x_2$.

 Let us discuss this case more precisely, focussing on $\alpha = 1/2$, for which the loci of the extrema can be found explicitly. The distribution $P(\omega)$ for this particular case is depicted in Fig.~\ref{sketch3}. One finds that the critical value of $\delta$ at which two inflection points emerge is
 \begin{equation}
 \delta_c = \left(17 + 12 \sqrt{2}\right)^{-1} \approx 0.0294.
 \end{equation}
For $\delta < \delta_c$, each extremum splits into a minimum and a maximum at locations
\begin{eqnarray}
\omega_{max,1} &=& 1 - \omega_{max,2} = \frac{1 + 7 \delta - \sqrt{1 - 34 \delta + \delta^2}}{8 (1 + \delta)}, \nonumber\\
\omega_{min,1} &=& 1 - \omega_{min,2} = \frac{1 + 7 \delta + \sqrt{1 - 34 \delta + \delta^2}}{8 (1 + \delta)}.
\end{eqnarray}
Further on, at $\delta = \delta_{cc}$, which is given by
\begin{equation}
\delta_{cc} = \frac{1}{11} \left(259 + 144 \sqrt{3} - 12 \sqrt{897 + 518 \sqrt{3}}\right) \approx 0.0108,
\end{equation}
all three maxima become equally high, \ie, $P(\omega = 1/2) = P(\omega_{max,1})= P(\omega_{max,2})$ (see the bold black line in Fig.~\ref{sketch3}). For yet smaller values of $\delta$, the maximum at $\omega = 1/2$ becomes smaller than the ones at $\omega = \omega_{max,1}$ and $\omega = \omega_{max,2}$. Hence, in the latter case it is most probable that $x_1$ and $x_2$ are disproportionately different and either of them, with equal probability, dominates the total wealth.

\section{The distribution $P(\omega)$ for an exponentially-tempered  Pareto law}

Consider next the exponentially-truncated $\Psi(x)$ in Eq.~(\ref{trunc2}). Substituting $\Psi(x)$ in Eq.~(\ref{trunc2}) into Eq.~(\ref{dist}),
we get
\begin{eqnarray}
\label{9}
 P(\omega) &=& \frac{\left(L H\right)^{\alpha}}{4 K_{\alpha}^2(2 \sqrt{\delta})} \, \frac{1}{\omega^{1+ \alpha} (1 - \omega)^{1 - \alpha}} \, \int^{\infty}_0 \frac{dx}{x^{1 + 2 \alpha}} \exp\left(- \frac{L}{\omega x} - \frac{x}{(1-\omega) H}\right) \nonumber\\
&=&
 \frac{1}{2 K_{\alpha}^2(2 \sqrt{\delta})} \frac{K_{2 \alpha}\left(2 \sqrt{\delta/\omega (1 - \omega)}\right)}{\omega (1 - \omega)} \,.
\end{eqnarray}
Note that $P(\omega)$ vanishes at the edges $\omega = 0$ and $\omega = 1$ exponentially fast and is symmetric around $\omega = 1/2$. Furthermore, one may readily check that the first derivative of $P(\omega)$ at $\omega = 1/2$ vanishes so that $P(\omega)$ is smooth at $\omega = 1/2$, as compared to the cusp-like behavior observed in case of a bounded PD. The question now is whether $\omega = 1/2$ is always a maximum.

\begin{figure}[ht]
  \centerline{\includegraphics[width=0.65\textwidth]{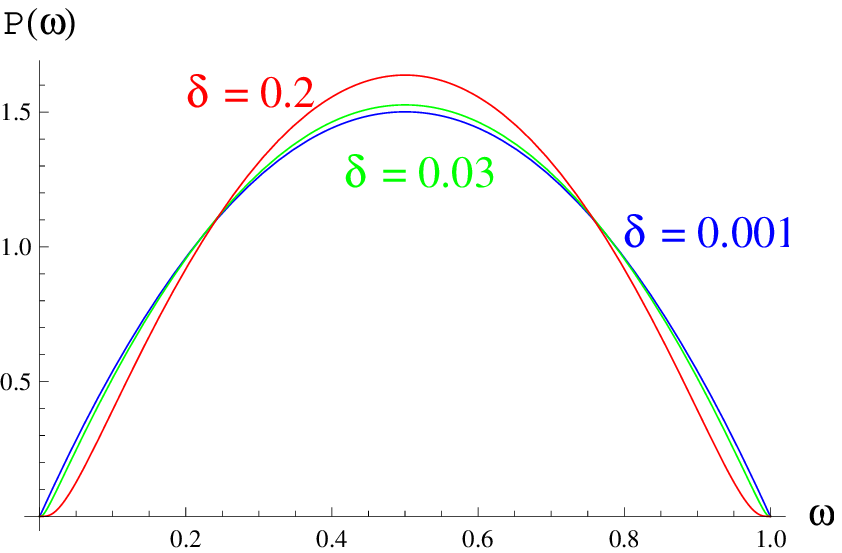}}
\caption{$P(\omega)$ in Eq.~(\ref{9}) for $\alpha = 2$ and  different $\delta$.
}
  \label{sketch4}
\end{figure}

Expanding $P(\omega)$ in Eq.~(\ref{9}) in a Taylor series around $\omega = 1/2$, we
get, omitting insignificant
numerical factors,
\begin{eqnarray}
\label{exp}
P(\omega) \sim 1 + g \, \left(\omega - \frac{1}{2}\right)^2 + {\mathcal O}\left((\omega - \frac{1}{2})^4\right), \, 
\end{eqnarray}
where the constant before the quadratic term reads
\begin{eqnarray}
g \sim 1 - \alpha - 2 \sqrt{\delta} \frac{K_{2 \alpha - 1}(4 \sqrt{\delta})}{K_{2 \alpha}(4 \sqrt{\delta})} \, . \label{g}
\end{eqnarray}

Observe that $g$ is always negative for any $\alpha \geq 1$. This means that
$P(\omega)$ is always a bell-shaped function with a \textit{maximum} at $\omega = 1/2$. Hence, as in  the case of a bounded PD, in this case it is most likely that the
wealths $x_1$ and $x_2$ of two randomly selected individuals will be the same. Typical profiles of $P(\omega)$ for $\alpha = 2$ are depicted in Fig.~\ref{sketch4} and show that the form of the distribution is not very sensitive to the value of $\delta$.

\begin{figure}[ht]
  \centerline{\includegraphics[width=0.65\textwidth]{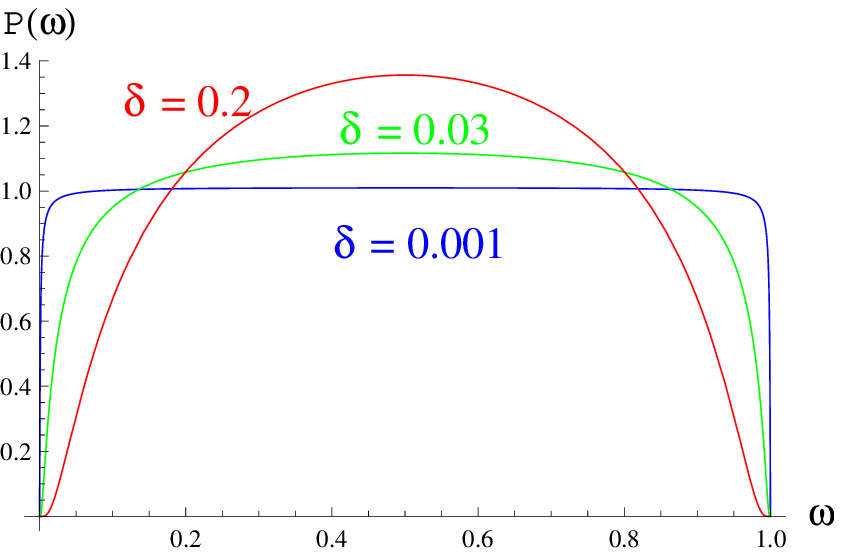}}
\caption{$P(\omega)$ in Eq.~(\ref{9}) for $\alpha = 1$ and  different $\delta$.
}
  \label{sketch5}
\end{figure}

Note next that the case $\alpha = 1$ appears to be somewhat special since $g \to 0$ when $\delta \to 0$. This means that for sufficiently small values of $\delta$ the distribution $P(\omega)$ will become nearly uniform (apart of an exponential truncation in the vicinity of the edges). This trend is clearly seen in Fig.~\ref{sketch5} and signifies that for such Paretian ensembles the contribution of each of the individuals to the total wealth can have \textit{any} value (inside the edges of the distribution) with equal probability.

\begin{figure}[ht]
  \centerline{\includegraphics[width=0.65\textwidth]{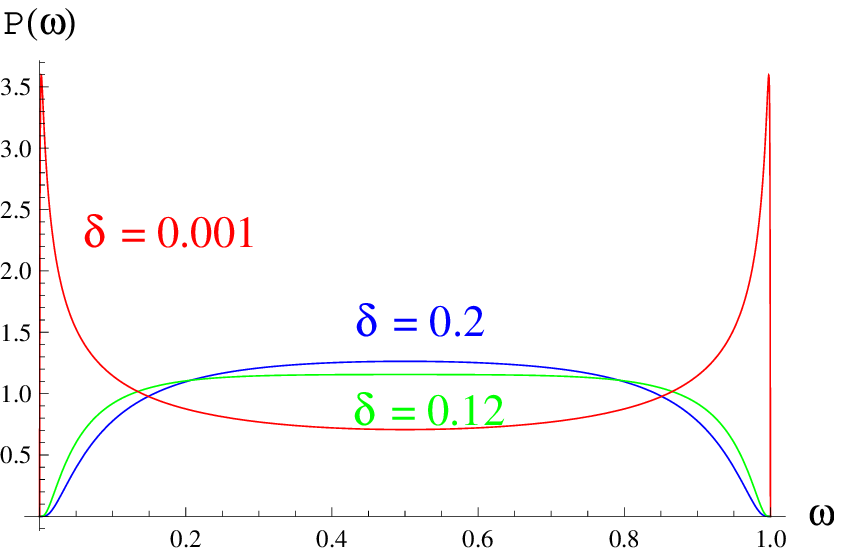}}
\caption{$P(\omega)$ in Eq.~(\ref{9}) for $\alpha = 1/2$ and  different $\delta$.
}
  \label{sketch6}
\end{figure}

We finally turn to the case $0 < \alpha < 1$. One can readily see that here, as in the
case of a bounded PD,
there exists a critical value $\delta_c$, defined implicitly by the following transcendental equation,
\begin{equation}
1 - \alpha = 2 \sqrt{\delta_c} \frac{K_{2 \alpha - 1}(4 \sqrt{\delta_c})}{K_{2 \alpha}(4 \sqrt{\delta_c})}.
\end{equation}
For $\delta > \delta_c$, $P(\omega)$ is always
a bell-shaped function with  a maximum at $\omega = 1/2$.
For
$\delta = \delta_c$,  $P(\omega) \approx 1$ except for narrow regions
at the edges, where it vanishes exponentially. Lastly,
for $\delta < \delta_c$,  $P(\omega)$ has
a bimodal $M$-like shape, with maxima close to $\omega = 0$ and $\omega = 1$,
$\omega = 1/2$ being the least probable value, which is different from the three-modal $W$-shaped form observed for a bounded PD.
This signifies again that for $\delta < \delta_c$ and $\alpha < 1$ the least probable event is that $x_1 = x_2$ and most likely either of two contributions completely dominate the total wealth, \ie, the distribution of wealth is disproportionate.
In Fig.~\ref{sketch6} we depict different possible forms of $P(\omega)$ for the particular case $\alpha=1/2$. Note that for an exponentially-truncated distribution the critical value of $\delta$ is somewhat larger, $\delta_c \approx 0.12$, than the corresponding value $\delta_c \approx 0.0294$ found for the bounded PD.

\section{Conclusions}

\begin{figure}[ht]
  \centerline{\includegraphics[width=0.65\textwidth]{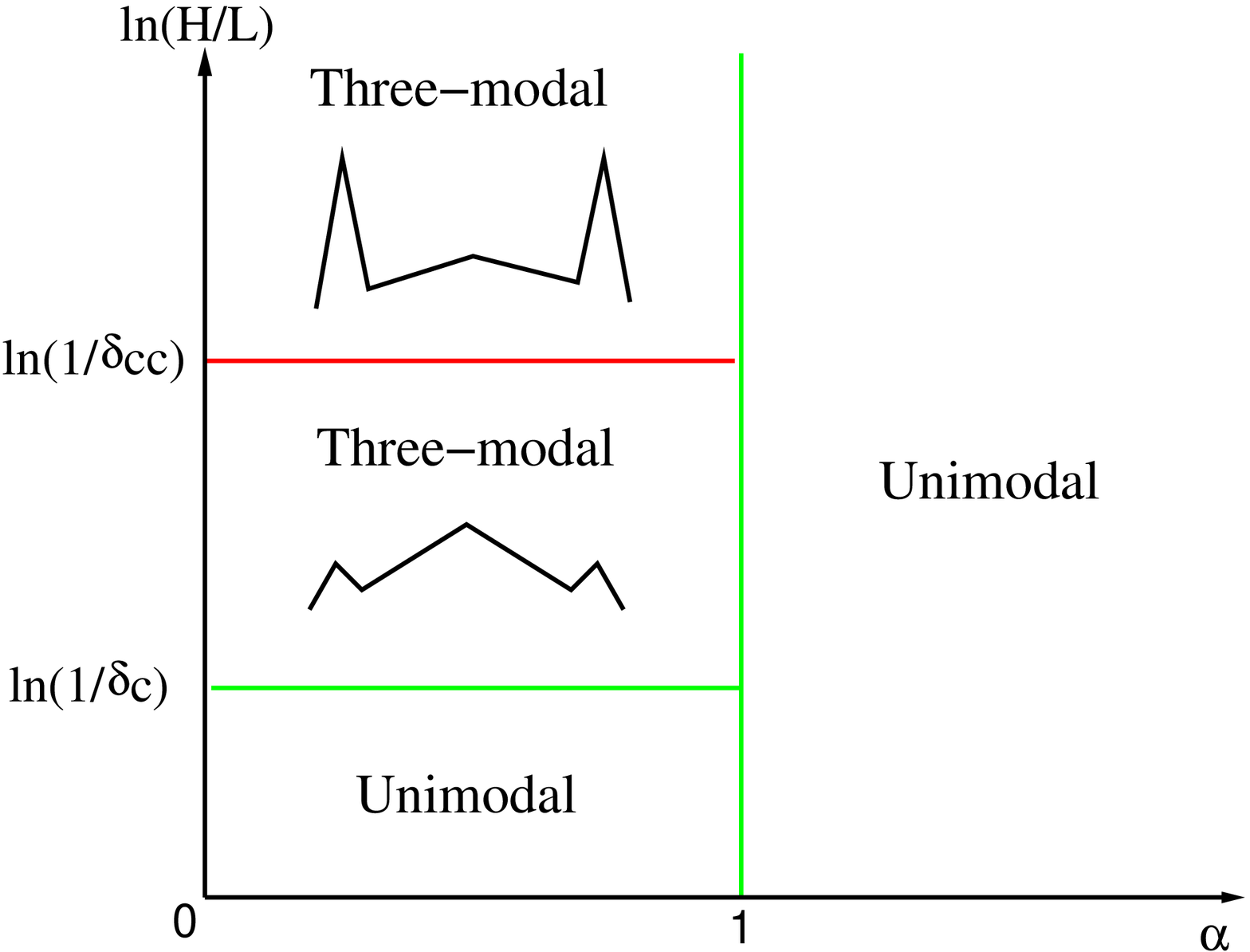}}
\caption{A "phase-diagram" for the bounded Pareto law.
A vertical line $\alpha = 1$ separates the domains in which $P(\omega)$ is always unimodal ($\alpha > 1$) or may have a different shape depending on the value of $\delta$. A horizontal line $\ln(\delta_c)$ separates the domain in which $P(\omega)$ is unimodal and the domain in which $P(\omega)$ attains a three-modal, $W$-like shape with the maximum at $\omega = 1/2$ being higher than two local maxima near the edges. Above the  horizontal line $\ln(\delta_{cc})$ one observes a  three-modal distribution $P(\omega)$ in which two maxima near the edges are higher than the one at $\omega = 1/2$.
}
  \label{sketch7}
\end{figure}

\begin{figure}[ht]
  \centerline{\includegraphics[width=0.65\textwidth]{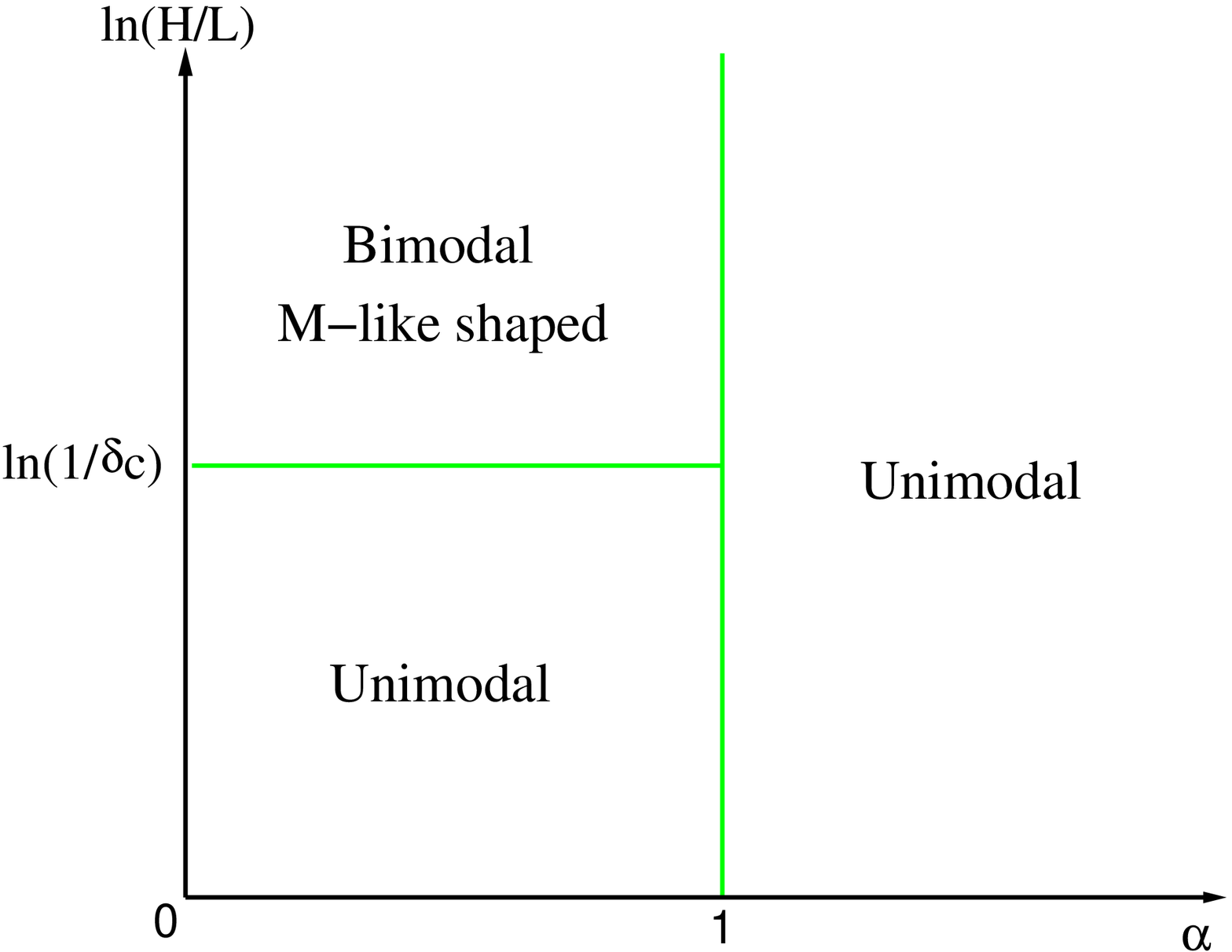}}
\caption{A "phase-diagram" for an exponentially-tempered  Pareto law. A vertical line $\alpha = 1$ separates the regimes in which $P(\omega)$ is always unimodal ($\alpha > 1$) or may have a different shape depending on the value of $\delta$. A horizontal line $\ln(\delta_c)$ separates the domain in which $P(\omega)$ is unimodal and the domain in which $P(\omega)$ attains a bimodal, $M$-like shape with a minimum at $\omega = 1/2$ and two maxima close to the edges.
}
  \label{sketch8}
\end{figure}

To conclude, we have studied here the distribution $P(\omega)$ of the random variable $\omega = x_1/(x_1 + x_2)$, where $x_1$ and $x_2$ are  the wealths of two individuals selected at random from
 the same tempered Paretian ensemble
 characterized by truncated Pareto distributions in Eqs.~(\ref{trunc1}) or (\ref{trunc2}).
 We have shown that, for $\alpha > 1$,  $P(\omega)$
always has a unimodal form and is peaked at $\omega = 1/2$, so that
most probably $x_1 \approx x_2$.
For $0 < \alpha < 1$ (which may be observed for developing economies, such as, \eg, India \cite{sinha}) we have encountered a more complicated behavior which
depends on the value of $\delta = L/H$.
In particular, we have realized that for
$\delta < \delta_c$ - a certain threshold value - $P(\omega)$ has a three-modal (for a bounded $\phi(x)$) and a bimodal $M$-shape (for an exponential $\phi(x)$) form which signifies
that in such ensembles the wealths $x_1$ and $x_2$ are disproportionately different.
Such a behavior appears to be quite surprising in view of the fact that the parental
 distributions $\Psi(x)$ have moments of arbitrary order. Our findings are summarized in the "phase-diagrams" presented in
 Figs.~(\ref{sketch7}) and (\ref{sketch8}).

We finally remark that a similar shape reversal of the distribution function has been observed for melting kinetics
of a heteropolymer \cite{redner} and for the Black-Scholes model of the stock options evolution in mathematical finance \cite{we}. Both works dealt with the parental distribution $\Psi(x)$ of the $x$-variables which, in addition to a power-law intermediate tail,
 has a log-normal truncation for large values of $x$, rather than an exponential one, and an exponential one for small values of $x$. This, of course, yields quantitatively
 different values of the critical parameters but qualitatively the effect is the same.
 We believe that the effect that we have found here is quite universal
for the tempered Paretian ensembles, and also quite robust. In particular,
we expect, as it was shown in Ref.~\cite{we} for different cut-off functions,
that it withstands correlations between $x_1$ and $x_2$, as long as the
correlation length does not exceed a certain critical value.

\section*{Acknowledgments}

We wish to thank I. Eliazar, A. Chattopadhyay and K. Lindenberg
for helpful comments.
Partial
financial support from the EU IRSES project N269139  "DCP PhysBio" for G.O. and Yu.H. is gratefully acknowledged.


\end{document}